\documentclass[
    prl,
    twoside,
    twocolumn,
    aps,
    10pt,
    floatfix,
    showpacs,
    citeautoscript,
    superscriptaddress,
]{revtex4-2}
\usepackage{amsmath,amssymb}
\usepackage{booktabs}
\usepackage{comment}
\usepackage{glossaries}
\usepackage{graphicx}
\usepackage[version=4]{mhchem}
\usepackage{siunitx}
\usepackage{tabularx}
\usepackage[dvipsnames]{xcolor}
\usepackage{hyperref}

\hypersetup{
    pdffitwindow=false,
    pdfstartview={FitH},
    pdfnewwindow=true,
    colorlinks=true,
    linkcolor=Black,
    citecolor=Black,
    urlcolor=Black,
}

\hypersetup{pdfauthor={S. Dutta, E. Fransson, T. hainer, B. M. Gallant, D. J. Kubicki, P. Erhart, and J. Wiktor}}
\hypersetup{pdftitle={Revealing the Low Temperature Phase of \ce{FAPbI3} Using A Machine-Learned Potential}}
\newcolumntype{P}[1]{>{\centering\arraybackslash}p{#1}}

\graphicspath{{figures/}}

\setacronymstyle{long-short}
\newacronym{acf}{ACF}{autocorrelation function}
\newacronym{dft}{DFT}{density functional theory}
\newacronym{gs}{GS}{ground state}
\newacronym{mas}{MAS}{magic angle spinning}
\newacronym{md}{MD}{molecular dynamics}
\newacronym{nmr}{NMR}{nuclear magnetic resonance}
\newacronym{nep}{NEP}{neuroevolution potential}

\DeclareSIUnit\angstrom{\text{Å}}
\DeclareSIUnit\atom{\text{atom}}

\renewcommand{\epsilon}[0]{\varepsilon}

\newcommand{\addchalmers}{
    Department of Physics,
    Chalmers University of Technology,
    SE-41296, Gothenburg, Sweden
}
\newcommand{\addbirmingham}{
    School of Chamistry,
    University of Birmingham,
    Edgbaston, B15 2TT,
    United Kingdom
}

\begin{document}

\title{
    Revealing the Low Temperature Phase of \texorpdfstring{\ce{FAPbI3}}{FAPbI3} using A Machine-Learned Potential
}

\author{Sangita Dutta}
\email{sangita.dutta@chalmers.se}
\author{Erik Fransson}
\author{Tobias Hainer}
\affiliation{\addchalmers}
\author{Benjamin M. Gallant}
\author{Dominik J. Kubicki}
\affiliation{\addbirmingham}
\author{Paul Erhart}
\author{Julia Wiktor}
\email{julia.wiktor@chalmers.se}
\affiliation{\addchalmers}


\begin{abstract}
\ce{FAPbI3} is a material of interest for its potential in solar cell applications, driven by its remarkable optoelectronic properties. However, the low-temperature phase of \ce{FAPbI3} remains poorly understood, with open questions surrounding its crystal structure, octahedral tilting, and the arrangement of formamidinium (FA) cations. Using our trained machine-learned potential in combination with large-scale molecular dynamics simulations, we provide a detailed investigation of this phase, uncovering its structural characteristics and dynamical behavior. Our analysis reveals the octahedral tilt pattern and sheds light on the rotational dynamics of FA cations in the low temperature phase. Strikingly, we find that the FA cations become frozen in a metastable configuration, unable to reach the thermodynamic ground state. By comparing our simulated results with experimental nuclear magnetic resonance (NMR) and inelastic neutron scattering (INS) spectra, we demonstrate good agreement, further validating our findings. This phenomenon mirrors experimental observations and offers a compelling explanation for the experimental challenges in accessing the true ground state. These findings provide critical insights into the fundamental physics of \ce{FAPbI3} and its low-temperature behavior, advancing our understanding of this technologically important material.
\end{abstract}

\maketitle

\section{Introduction}

Perovskite solar cells are recognized as promising optoelectronic devices due to their band gap favorably matching the solar spectrum \cite{PSC, PVC1, C1NR10867K, Frost2014, Stoumpos2013}.
Among various materials, hybrid halide perovskites, particularly methylammonium lead iodide (\ce{MAPbI3}) and formamidinium lead iodide (\ce{FAPbI3}), have attracted significant attention for next-generation photovoltaics.
Their efficiency has rapidly increased beyond \qty{25}{\percent} since their initial application \cite{PVC1, Tuo, Stoumpos2013}.
However, stability issues remain a major limitation, driving research into their crystal structure dynamics and phase stability \cite{PhysRevLett.122.225701, stability, C5EE03255E, C6CP02917E, Fabini2017}. 
Previous studies have highlighted the crucial role of rotational dynamics of organic cations and octahedral tilting in hybrid halide perovskites, influencing not only phase stability but also carrier lifetimes and overall device performance.
Neglecting these dynamics can lead to misinterpretations in experimental studies, particularly for techniques sensitive to local structural variations \cite{druzbicki_cation_2021, Lavén2023, FraRosEri23, WikFraKub23, Adams:ct5020}.

\ce{FAPbI3} has emerged as a preferred choice for photovoltaic thin films due to superior optoelectronic properties \cite{Tuo}.
At room temperature, it adopts a cubic structure, transitioning to the tetragonal $\beta$-phase below \qty{285}{\kelvin}, and further to the $\gamma$-phase at \qty{150}{\kelvin} \cite{Fabini2017, Tuo, Charles, Stoumpos2013, Oliver}.
Notably, ambiguity persists regarding the nature of the low-temperature $\gamma$-phase, with several experimental studies suggesting possible structural disorder \cite{Fabini2017, Tuo, Lavén2023, Charles, Stoumpos2013, Oliver}.
However, the exact nature of this disorder remains unresolved.
In this work, given the need for a detailed understanding of the low-temperature crystal structure and FA dynamics, we employ atomic-scale simulations to investigate the microscopic behavior of the $\gamma$-phase.

Computational studies of halide perovskite structures face challenges due to the strong anharmonicity of these materials and the rotational degrees of freedom of the organic cations.
Conventional static calculations provide limited insight while perturbative approaches are hindered by the strong anharmonicity, necessitating \gls{md} simulations to capture finite-temperature effects.
However, ab-initio \gls{md} simulations are computationally expensive, restricting access to long timescales and large system sizes.
Recently, machine-learned interatomic potentials have emerged as powerful tools for studying halide perovskite dynamics, enabling efficient sampling without compromising accuracy \cite{PhysRevLett.122.225701, Fransson2023, Fransson20231, Wiktor2023, Baldwin2024, Zhou2020, Bokdam2021}.

Here, we employ a machine-learned interatomic potential recently developed for the \ce{MA_{1-x}FA_xPbI3} system \cite{hainer_2025}, based on the fourth-generation \gls{nep} framework \cite{nep3, SonZhaLiu24}, to analyze the atomic scale dynamics of \ce{FAPbI3} via \gls{md} simulations.
Notably, the machine-learned potential accurately reproduces all known phases of \ce{FAPbI3} reported in the literature \cite{Stoumpos2013, Fabini2017, Charles, Oliver, Tuo}.
We first identify the ground-state structure as $a^-b^-b^-$ in Glazer notation \cite{Glazer:a09401}.
We then analyze octahedral tilting and FA molecular orientation across different phases.

Our simulations reveal that the low-temperature phase exhibits an $a^-a^-c^+$ structure due to kinetic trapping in a metastable state during cooling.
To understand this phenomenon, we further investigate the complex dynamics of organic cations, their correlations, and the associated free energy landscape.

\section{Methods}

\subsection{MD simulations}
\Gls{md} simulations were carried out using the \textsc{gpumd} package with a time-step of \qty{0.5}{\femto\second}.
We use a \gls{nep} trained for a mixed \ce{FA_{1-x}MA_xPbI3} system as described in Ref.~\citenum{hainer_2025}.
The potential was trained against \gls{dft} data generated using the SCAN+rVV10 functional \cite{peng_versatile_2016}.
The reference data comprised a wide range of configurations representing both \ce{FAPbI3}, \ce{MAPbI3}, and mixtures thereof.
The model, as well as the training data, are available on zenodo (https://doi.org/10.5281/zenodo.14992798).
We employed the Bussi-Donadio-Parrinello thermostat \cite{BDP} and the stochastic cell rescaling (SCR)
barostat \cite{Bernetti} method to control the temperature and pressure, respectively.
A system of \num{49152} atoms was chosen to avoid finite size effects \cite{Erk_finite_size_effects}.
We ran heating and cooling \gls{md} simulations in the NPT ensemble within \num{0} to \qty{350}{\kelvin} temperature span with different heating and cooling rates. Further details on the MD analyses, including structural and dynamical characterizations, are presented in the Results section.

\subsection{NMR measurements}

In order to determine the local environment of FA in the $\gamma$-phase of \ce{FAPbI3}, we carried out low-temperature \gls{mas} solid-state $^{13}$C and $^{15}$N \gls{nmr} measurements on single crystals of 3D perovskite \ce{FAPbI3}.

FAPbI$_3$ single crystals were fabricated following a previously published protocol~\cite{duijnstee2023understanding}. Briefly, a 1~M solution of formamidinium iodide (687.9~mg, 4~mmol; $>$99.99\%, Greatcell Solar Materials) and lead(II) iodide (1844.0~mg, 4~mmol; 99.99\% trace metal basis, Tokyo Chemical Industries) in 4~mL $\gamma$-butyrolactone (Alfa Aesar) was prepared. The solution was stirred at 60~°C for 4 hours, then filtered with a 25~mm diameter, 0.45~\textmu m pore glass microfibre filter. The filtrate was placed in a vial and heated in an oil bath undisturbed at 95~°C for 4 hours until small crystals formed. The crystals were then dried in a vacuum oven at 180~°C for 45 minutes. All synthetic work besides drying was conducted in an N$_2$ glovebox.

MAS NMR spectroscopy was carried out using a commercial Bruker Avance Neo 400~MHz spectrometer equipped with an LTMAS 3.2~mm Bruker $^1$H/X/Y triple-resonance probe. All measurements were conducted at approximately 95~K using an 8~kHz MAS spin rate. For both $^{13}$C and $^{15}$N measurements, a $^1$H-X cross-polarisation (CP) MAS pulse sequence was used. $\gamma$-Glycine was used to calibrate the $^1$H, $^{13}$C, and $^{15}$N radiofrequency field amplitudes (60, 40, and 140~kHz, respectively) and CP contact times (1~ms and 3~ms for $^1$H-$^{13}$C and $^1$H-$^{15}$N, respectively), and to reference $^{13}$C and $^{15}$N chemical shifts (174.9~ppm for $^{13}$C of C=O; 32.9~ppm for $^{15}$N). $^1$H decoupling at an RF field of 60~kHz was used during acquisition in all measurements. We summarise the experimental parameters for all NMR measurements reported here in \autoref{stab:nmr}.

Immediately prior to measurement, the crystals were gently crushed and heated at 150~°C on a hot plate to ensure they were in the 3D FAPbI$_3$ $\alpha$-phase. These crushed crystals were packed inside a 3.2~mm sapphire rotor. The same packed rotor was used for all measurements reported here. The crystals were rapidly cooled (\textit{freeze}) from 298~K to 95~K at a rate of 5000--10000~K~min$^{-1}$ by inserting the rotor into the probe at 95~K. Between each measurement, the crystals were rapidly warmed to 298~K by ejecting the rotor into ambient air, where it was kept for at least 5~minutes before the next cooling cycle. Notably, prior to the first measurements (freeze~1) the rotor had been cooled and heated in this manner several times. We therefore discount a difference between the first and subsequent quenching events as the source of observed $^{15}$N spectral differences between freeze~1, freeze~2, and freeze~3.

\subsection{Calculation of \texorpdfstring{$^{15}$}{15}N Chemical Shifts}

First-principles calculations of $^{15}$N chemical shifts were performed using \gls{dft} with the \textsc{Quantum ESPRESSO} \cite{giannozzi2009quantum, giannozzi2017advanced} package, employing the Perdew-Burke-Ernzerhof exchange-correlation functional and the gauge-including projected augmented wave method \cite{pickard2001all, charpentier2011paw}. 

Calculations were performed for two types of structures: the ground-state $a^-b^-b^-$ structure and three representative configurations of the cooled $a^-a^-c^+$ structure.
In the latter case, atomic configurations for shielding calculations were extracted from molecular dynamics cooling simulations conducted in a 96-atom supercell.
We set the plane-wave energy cutoff of \qty{80}{Ry} for wavefunctions and \qty{640}{Ry} for the charge density.
We used a $\Gamma$-centered \numproduct{2x2x2} $k$-point grid for Brillouin zone sampling.  

To relate the computed trace of the shielding tensor $\sigma_\text{calc}$ to experimental $^{15}$N chemical shifts $\delta_\text{exp}$, an empirical scaling was applied based on reference data \cite{hartman2016benchmark}.
The scaling was performed via linear regression of computed shieldings against experimentally measured chemical shifts from LGLUAC11, GLUTAM01, BITZAF, and CIMETD.
This set corresponds to ten inequivalent local environments for N, spanning chemical shifts from \num{-1.3} to \qty{249.5}{ppm}.

The final chemical shifts were obtained using the linear transformation:
\begin{align}
    \delta_\text{calc} = a \cdot \sigma_\text{calc} + b,
\end{align}
where the parameters $a$ and $b$ of \num{-1.05} and \num{201.88}, respectively, were determined empirically from regression analysis of the reference dataset.

\subsection{Dynamical structure factor from MD}

We compute the dynamical structure factor from \gls{md} simulations using the \textsc{dynasor} package \cite{FraSlaErhWah2021}.
For each structure prototype, we run 40 independent simulations, each \qty{100}{\pico\second} long and average $S(q, \omega)$ over all the runs.
The total $S(q, \omega)$ is given by the sum of the coherent and incoherent dynamical structure factors which are weighted with their respective neutron scattering lengths.
The resulting vibrational spectra are dominated by hydrogen motion due to its large incoherent scattering length.
Since hydrogen dynamics is mostly $q$ independent, we sum $S(q, \omega)$ over $q$-points between 0 and \qty{15}{rad/\angstrom}.
The spectrum is calculated at \qty{10}{\kelvin}, which means that the classical spectra obtained from \gls{md} does not capture the correct quantum statistics (intensities of the peaks).
Therefore, we rescale the spectrum by
\begin{equation}
    S_{QM}(q,\omega) = \frac{\omega}{1-\exp{(\hbar \omega / k_\text{B}T})} S(q,\omega)
\end{equation}
as described in Ref.~\cite{RosFraOst25}.

\begin{figure}
    \centering
    \includegraphics{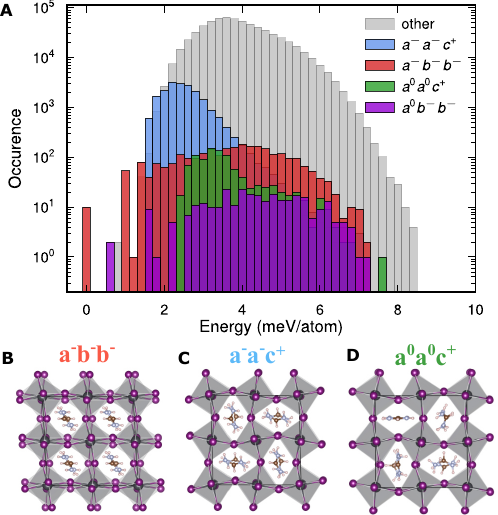}
    \caption{
        (A) Energy distribution of fully relaxed perovskite phases of \ce{FAPbI3} structures obtained by considering \num{1000} of different tilted structures with randomized FA orientations in \numproduct{2x2x2} supercells of a corresponding primitive cell.
        Relevant low energy structures are marked with color.
        Structural view of (B) $a^-b^-b^-$, (C) $a^-a^-c^+$, (D) $a^0a^0c^+$ phases are shown.
    }
    \label{fig:ground_state}
\end{figure}

\section{Results}

\begin{figure}
    \centering
    \includegraphics{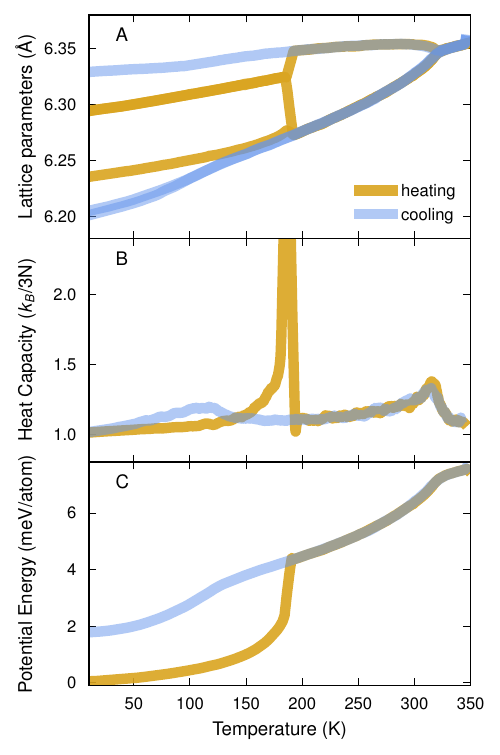}
    \caption{
        (A) Lattice parameters, (B) heat capacity, (C) Energy from heating and cooling \gls{md} with 6.34~K/ns rate, respectively in \ce{FAPbI3}.
    }
    \label{fig:lat_heatcap_data}
\end{figure}

\subsection{Searching for the Lowest energy structure in \texorpdfstring{\ce{FAPbI3}}{FAPbI3}}

To understand the energy landscape of \ce{FAPbI3}, we perform an extensive sampling of possible structures as shown in \autoref{fig:ground_state}.
About a million initial structures are created in \numproduct{2x2x2} supercells of the cubic primitive cell, incorporating randomized FA orientations and tilt modes with random mode amplitudes for each Cartesian direction.
We relax each structure until the largest force on any atom falls below \qty{0.1}{\milli\electronvolt\per\angstrom}.
The resulting perovskite structures are then classified into Glazer structures \cite{Glazer:a09401} by projection onto the M and R phonon modes (corresponding to octahedral tilting) as done in Refs.~\citenum{FraRahWik23, hainer_2025, Kayastha2025}.
The \gls{gs} perovskite structure is identified as $a^-b^-b^-$ in the Glazer space as indicated in red in \autoref{fig:ground_state}A.
\autoref{fig:ground_state}B shows the structure of $a^-b^-b^-$ where all FAs are pointing in the same direction.
The second lowest energy structure is identified as $a^0b^-b^-$, which is structurally very similar to the ground-state but lacks a small out-of-phase tilt around the $x$-axis.
We also identify other possible structures with small energy differences, competing with the \gls{gs} structure seen in \autoref{fig:ground_state}A.
The atomic structures with preferred FA orientations of other relevant low energy structures, i.e., $a^-a^-c^+$ and $a^0a^0c^+$ are shown in \autoref{fig:ground_state}C and D, respectively.
The total energies calculated using \gls{nep} and \gls{dft} are provided in the \autoref{stab:energy}, demonstrating good agreement with \gls{dft} calculations.

\begin{figure}
    \centering
    \includegraphics{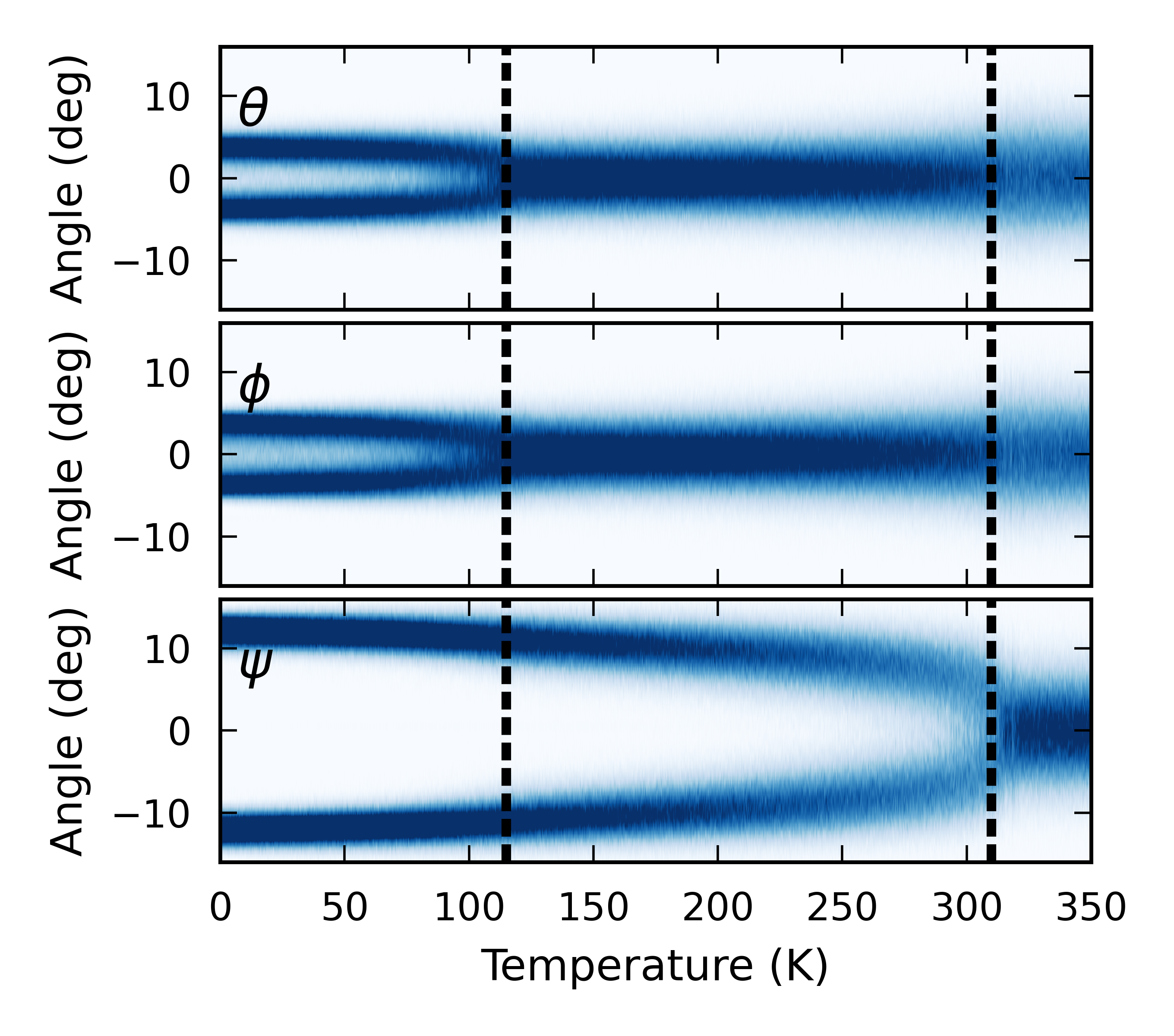}
    \caption{
        Maps of tilt angles as a function of temperature from cooling \gls{md} runs.
        Dashed white lines represent two successive phase transitions from $a^0a^0a^0$ to $a^0a^0c^+$ and $a^0a^0c^+$ to $a^-a^-c^+$-phase.
    }
    \label{fig:tilt_data}
\end{figure}

\subsection{Behavior during Cooling and Heating}
    
After identifying the most stable structure at \qty{0}{\kelvin}, we now perform heating and cooling runs to assess the phase transitions and compare them with experimental findings.
Phase transitions can readily be seen as discrete or continuous changes in the thermodynamic properties like energy, heat-capacity and lattice parameters.
To check the rate effects, we run simulations with different heating and cooling rates considering a supercell which is equivalent to a \numproduct{16x16x16} primitive cubic (12-atom) cell, and \numproduct{8x8x8} $a^-b^-b^-$ (96-atom) cell.
The convergence of the lattice parameter, energy, and heat capacity with respect to the heating and cooling rate can be found in \autoref{sfig:diff_rate}.

\autoref{fig:lat_heatcap_data} shows the mentioned parameters as a function of temperature with the slowest heating and cooling rate (\qty{6.34}{\kelvin\per\nano\second}).
On heating, starting from the $a^-b^-b^-$ structure, the simulation yield a transition to $\beta$-phase at about \qty{190}{\kelvin} and then to $\alpha$-phase at about \qty{315}{\kelvin}.
In the cooling run the simulation captures the same $\alpha$ to $\beta$ transition, however, its transition into a different low temperature phase occurs at about \qty{120}{\kelvin}, which is \qty{2}{meV/atom} higher in energy than the ground-state.
The low temperature transition thus exhibits hysteresis, and in the heating run appears to be of first-order in character.
In contrast, the $\beta$ to $\alpha$-phase transition is a continuous one.

Here, it is interesting to note again that the low temperature structure obtained from cooling in experiments is not fully understood \cite{Fabini2017, Lavén2023, Tuo, Charles, Stoumpos2013, Oliver}.
To determine whether the structure found in our cooling simulations corresponds to the one encountered in experimental studies, we therefore analyze it in more detail.

\begin{figure}
    \centering
    \includegraphics{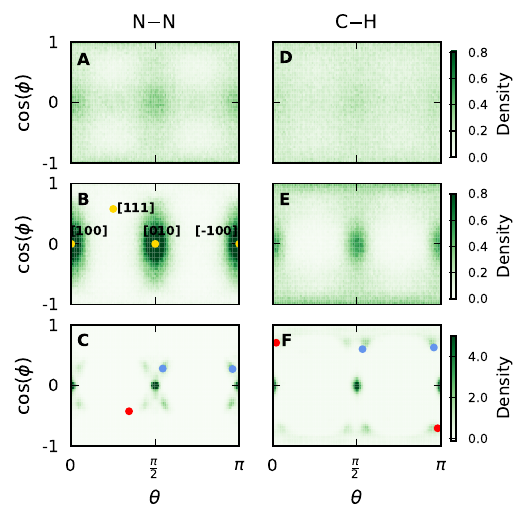}
    \caption{
        Probability distribution $P(\theta, \phi)$ of N--N vectors (A), (B), and (C), and of C--H vectors (D), (E), and (F) in $a^0a^0a^0$, $a^0a^0c^+$, and $a^-a^-c^+$ phases (top to bottom), respectively.
        This $a^-a^-c^+$ phase is obtained from the cooling run.
        Here, $\theta$ refers to angle in the $x-y$ plane and $\phi$ is angle to the $z axis$.
        Position of the vectors in Cartesian coordinates are marked in red color in (B).
        The orientation of the N--N and C--H vectors of FAs in an ideal $a^-a^-c^+$ phase and $a^-b^-b^-$ phase are shown by the blue and red dots in (C) and (F), respectively.
    }
    \label{fig:heatmap}
\end{figure}

\subsection{Tilt angle analysis} 
To gain additional insight into the low temperature phase obtained from the cooling run, we first focus on the octahedra tilting patterns of the system at different temperatures.
Here, we compute the \ce{PbI6} (see \autoref{sfig:octahedra}a) octahedral tilt angles in the perovskite structures during cooling \gls{md} simulations as done in Refs.~\citenum{WikFraKub23} and \citenum{Baldwin2023}.
First, the \ce{PbI6} octahedron is matched to a fully symmetric octahedron in an ideal cubic perovskite following the algorithm in Ref.~\citenum{Larsen_2016} as implemented in \textsc{ovito} \cite{Stukowski2010}, which generates the rotation and scales for optimal mapping.
Functionality from the \textsc{scipy} package \cite{Virtanen2020} is used to convert the rotation to Euler angles (see \autoref{sfig:octahedra}b for the definition of the Euler angles).
Following Glazer's approach \cite{Glazer:a09401}, we choose the rotation which produces the angles in increasing magnitude among the three possible options.

The distribution of octahedral tilt angles over the entire temperature from the cooling run is shown in \autoref{fig:tilt_data}.
The transition temperatures obtained from \autoref{fig:lat_heatcap_data}, are indicated by vertical dashed lines.
In the high-temperature $\alpha$-phase, which can be described as $a^0a^0a^0$ in Glazer notation, the tilt angle distributions are monomodal and centered around 0$^\circ$.
Next, in the $\beta$-phase, the $\psi$ angle, which characterizes the tilt in the $z$ direction, obtains an average value of about \qty{10}{\degree}, which upon visual inspection with \textsc{ovito} can be identified as an in-phase tilting pattern.
Glazer notation thus describes this $\beta$ phase as $a^0a^0c^+$. 
The tilt angles $\theta$ and $\phi$ become non zero in the low temperature $\gamma$ phase.
After analysis of tilt patterns in all directions, we found that the $c^+$ tilt from $a^0a^0c^+$ structure becomes more robust with an average value of about \qty{15}{\degree} in the $\gamma$-phase.
Additional out-of-phase tilt with a value of $\theta = \phi \simeq \qty{5}{\degree}$ appears along the $x$ and $y$ directions.
Thus, one can characterize this $\gamma$-phase as $a^-a^-c^+$ in the Glazer space. The snapshots obtained from the cooling simulation run, highlighting representative temperatures and corresponding octahedral tilt configurations, are shown in the \autoref{sfig:snapshot}.
It is important to note that for another similar FA-based perovskite \ce{FAPbBr3} structure below \qty{153}{\kelvin} also has been experimentally identified as the same $a^-a^-c^+$ ($Pnma$) phase \cite{Simenas2024}.

As noted earlier, the structure we find upon cooling does not correspond to the \gls{gs} structure of \ce{FAPbI3} identified in the previous section (\autoref{fig:lat_heatcap_data}).
This suggests two possibilities: (i) our \gls{md} simulations do not reach the true low-temperature structure of \ce{FAPbI3} ($a^-b^-b^-$) due to limitations in cooling rates, whereas experiments do, or (ii) the $a^-a^-c^+$ structure represents a frozen metastable state, mirroring a physical scenario where \ce{FAPbI3} remains kinetically trapped during cooling instead of transitioning to the \gls{gs} structure, which is also the case in experiments.
To test these hypotheses, we will analyze the ordering and dynamics of FA molecules and compare simulated characteristics of the potential phases to experimental measurements.

\subsection{Ordering of FAs}
To understand the local symmetry, we start looking at the molecule reorientation in different phases of \ce{FAPbI3}.
We consider the vector connecting the two N atoms, $\mathbf{r}_{\text{NN}}$, and the vector between C and H atoms, $\mathbf{r}_{\text{CH}}$, in a FA molecule as shown in \autoref{sfig:octahedra}c.
We compute the orientation represented by the polar angle $\phi$ and azimuthal angle $\theta$ for each of them.
$\phi$ is the angle between $\boldsymbol{r}_{\text{NN}}$ ($\boldsymbol{r}_{\text{CH}}$) and $z$ direction, and $\theta$ denotes the angle in the $xy$ plane.
\autoref{fig:heatmap} represents the probability distributions over $\theta$ and $\phi$ ($P(\theta,\phi)$) for N--N and C--H vectors in the three different phases of \ce{FAPbI3} from the cooling run.

In the high temperature $a^0a^0a^0$ phase (at \qty{330}{\kelvin}), the N--N and C--H vectors are homogeneously distributed, indicating an almost-free molecular rotation of FA molecules as shown in \autoref{fig:heatmap}A and D.
Once cooled down from the $a^0a^0a^0$ to the $a^0a^0c^+$-phase (at \qty{200}{\kelvin}), we notice a pattern appearing in the distributions, which is symmetric in the $xy$ plane as shown in \autoref{fig:heatmap}B and E, also observed by Tua \textit{et al.} \cite{Tuo}.
The N--N vectors are most likely to be aligned with the $x$ ([100]) and $y$ ([010]) directions. 
This arrangement of FA molecules in \autoref{fig:heatmap}B and E is also reflected in the \numproduct{2x2x2} supercell of $a^0a^0c^+$ structure.
The pattern is not as clear for C--H compared to N--N, however it shows some preferred orientations along [100], [010], and [001] directions.

The distribution becomes sharper and changes again when cooled down into the $a^-a^-c^+$ phase (at \qty{10}{\kelvin}) for both N--N and C--H vectors (\autoref{fig:heatmap}C, F).
The distributions mostly retain the preferred orientations from the $a^0a^0c^+$ phase but with four symmetric additional orientations appearing as ``wings''. These wings correspond to the orientations found in the ideal $a^-a^-c^+$ structure (shown in \autoref{fig:ground_state} C), as marked by the blue dots in \autoref{fig:heatmap}C, F. However, the cooled structure differs from significantly from the ideal one, as it still has a large proportion of FA molecules stuck in the orientations characteristic for the $a^0a^0c^+$ phase.
Note here that the four symmetric ``wings'' each correspond to a symmetrically equivalent version of the $a^-a^-c^+$ structure.

We also compare the FA orientations to that of the \gls{gs} phase, represented as red dots in \autoref{fig:heatmap}C and F. 

This orientation corresponds to a very low probability distribution at the temperatures close to the transition.
Therefore, a significant free energy barrier likely prevents the FA molecules from aligning as in the \gls{gs} phase, leading to structural freezing in a metastable state.
To have a quantitative picture, we estimate that the free energy barrier for N--N vectors to align like in the \gls{gs} phase using $F=-k_Bln(P(\theta,\phi))$, where $k_B$ is the Boltzmann constant, and find it to be more than \qty{100}{\milli\electronvolt} per FA at \qty{200}{\kelvin} (see \autoref{sfig:energy_barr}).

Next, we assess the ordering of FA molecules in the different relevant structures.

This is done by analyzing the nearest neighbor correlation of N--N and C--H vectors as shown in \autoref{sfig:neighbor} at different temperatures.
The results highlight that the $a^-a^-c^+$ structure found upon cooling is significantly more disordered than the ideal $a^-a^-c^+$ and $a^-b^-b^-$ phases.
Notably, the ideal $a^-a^-c^+$ phase loses its strong ordering (and approaches that of the cooled structure) when heated up to only \qty{50}{\kelvin}, whereas the ground-state, $a^-b^-b^-$, remains very ordered indicating FAs are more locked into place in this phase.

These analyses of the FA orientational distributions and ordering demonstrate that the cooled structure has several different local FA orientations and environments, indicating more disorder compared to the ideal structures.
This is qualitatively in agreement with experimental studies, which observe a significant degree of disorder in the low-temperature structure~\cite{Fabini2017, Oliver}.

\begin{figure}
    \centering
    \includegraphics{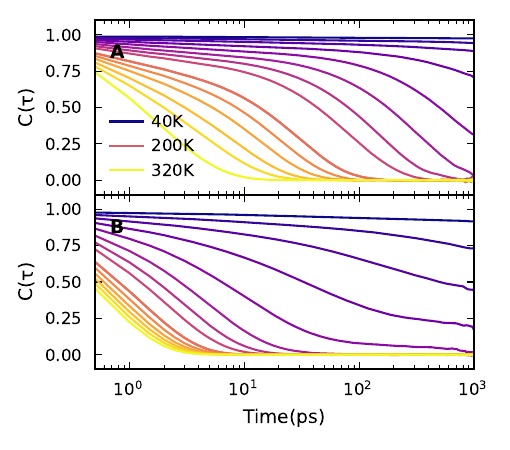}
    \caption{
        Autocorrelation function $C(\tau)$ for the orientation of (A) N--N and (B) C--H vector in FA units.
        The spacing between the line is \qty{20}{\kelvin}.
    }
    \label{fig:correlation}
\end{figure}

\subsection{Rotational dynamics of FAs}

Next, we analyze the rotational dynamics of FA molecules by calculating the orientational autocorrelation function (ACF) as defined in 
\begin{equation}
    C(\tau)=\frac{\left\langle{\mathbf{r}^{i}_{\text{NN}}(t)}{\mathbf{r}^{i}_{\text{NN}}(t+\tau)}\right\rangle}{\left\langle{\mathbf{r}^{i}_{\text{NN}}(t)}{\mathbf{r}^{i}_{\text{NN}}(t)}\right\rangle}
\end{equation}
where $\mathbf{r}^{i}_{\text{NN}}(t)$ ($\mathbf{r}^{i}_{\text{CH}}(t)$) is the N--N (C--H) bond vector at time $t$ for the $i$th FA molecule. To this end, we run \gls{md} simulations at several temperatures starting from the phase corresponding to those temperatures. The N--N (C--H) bond vector $\mathbf{r}_{\text{NN}}$ ($\mathbf{r}_{\text{CH}}$) of each FA unit is sampled in the NVE ensemble for \qty{1000}{\pico\second} (at the volume previously obtained from NPT runs).
\autoref{fig:correlation}A and B represent the \gls{acf} of the N--N, and C--H axis as a function of time, respectively.
The \gls{acf} decays faster at high temperature, reflecting faster reorientation of the FA molecules in the high temperature phase.
However, it decays more slowly with decreasing the temperature, indicating freezing of FA molecules.

\begin{figure}
    \centering
    \includegraphics{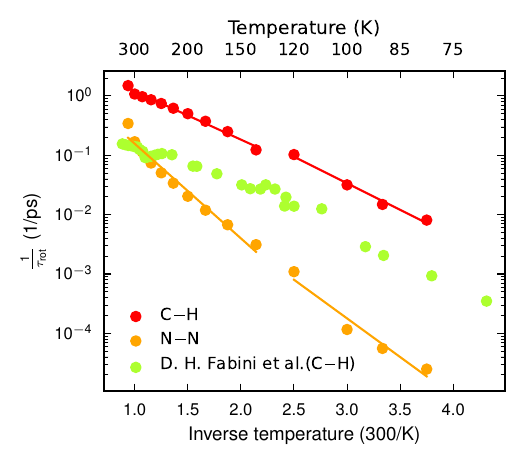}
    \caption{
        Rotation rate ($1/\tau_\text{rot}$) as a function of temperature.
        The solid lines correspond to Arrhenius fits mentioned in the text.
        Green symbols represent the data for rotation of C--H vectors from experiment for comparison \cite{Fabini2017}.
    }
    \label{fig:arrhenius}
\end{figure}

The decay in the \gls{acf} can be modeled with an exponential function as $C(\tau)\propto e^{-\tau/\tau_\text{rot}} + e^{-\tau/\tau_\text{vib}}$, where $\tau_\text{rot}$ denotes the rotational times, and $\tau_\text{vib}$ accounts for vibrations of of the FA molecule~\cite{EriF, FraRosEri23}.
The rotational times of the N--N and C--H vectors are shorter for the tetragonal $a^0a^0c^+$ phase (above \qty{120}{\kelvin}) than those of the low temperature phase (below \qty{120}{\kelvin}).
\autoref{fig:arrhenius} displays estimation of the rotational time of N--N and C--H vectors.
The rotational times of C--H axis measured in experiment \cite{Fabini2017} are in reasonable agreement with our predicted values.
The offset between the present study and experiment can possibly be attributed to the the model accuracy and difficulties in capturing slow dynamics of FA molecules in \gls{md}.

Subsequently, we model the temperature dependence of the rotational time using the Arrhenius equation, $1/\tau_\text{rot}\propto e^{-E_{\rm A}/k_BT}$, where $E_{\rm A}$ is the activation energy and $k_B$ is the Boltzmann constant, which fits the data well.
This yields the activation barrier of the rotational process for different phases, which are provided in \autoref{tab:activation} along with a comparison with literature.
We find good agreement with experimentally measured and calculated values from Ref.~\citenum{Fabini2017}.
Furthermore, we note that the barrier for the N--N vector in the $a^0a^0c^+$ phase, \qty{94.9}{meV}, is consistent with the barrier obtained from the free energy landscapes at \qty{200}{\kelvin} (\autoref{sfig:energy_barr}).

Lastly, we compare the dynamics of FAs obtained above with the \gls{gs} structure, \autoref{sfig:correlation_GS}.
Interestingly, the \gls{acf} in the \gls{gs} phase indicates that all the FA molecules are frozen with $C_i(\tau)\sim 1$ throughout the time range (\qty{10}{ns}) and up to \qty{120}{\kelvin}.
A rough estimate of the rotational time for the very flat \gls{acf} at \qty{120}{\kelvin} in the \gls{gs} is at least \qty{20}{\micro\second}.
This suggests that the FAs in this phase do not rotate, unlike in the experimentally observed low temperature phase where they rotate on a nanosecond time scale at these temperatures. This indicates that the experimental phase does not reach the ground-state structure and that the kinetic trapping observed in our simulations reflects a physically realistic metastable state.

\begin{table}
\centering
\caption{
    Activation energy barriers in \unit{\milli\electronvolt} for molecular rotation along the N--N and C--H axes from the present study and the available literature \cite{Fabini2017} for the $a^0a^0c^+$ and $a^-a^-c^+$ phases of \ce{FAPbI3}.
}
\begin{tabular}{lP{1.5cm}P{1.5cm}cP{1.5cm}P{1.5cm}}
    \toprule
    & \multicolumn{2}{c}{$\mathbf{a^0a^0c^+}$}
    &
    & \multicolumn{2}{c} {$\mathbf{a^-a^-c^+}$} \\ 
    \cline{2-3}
    \cline{5-6}
    \\[-6pt]
    &            C--H & N--N && C--H & N--N \\  
    Experiment~\cite{Fabini2017} & 45   &   -- && 84   &   -- \\
    \gls{dft} \cite{Fabini2017}       & 39   &   -- && 63   &   -- \\
    \gls{nep}    & 48.5 & 94.9 && 53.3 & 77.5 \\
    \bottomrule
\end{tabular}
\label{tab:activation}
\end{table}

\subsection{Experimental verification }
\begin{figure*}
    \centering
    \includegraphics[width=\textwidth]{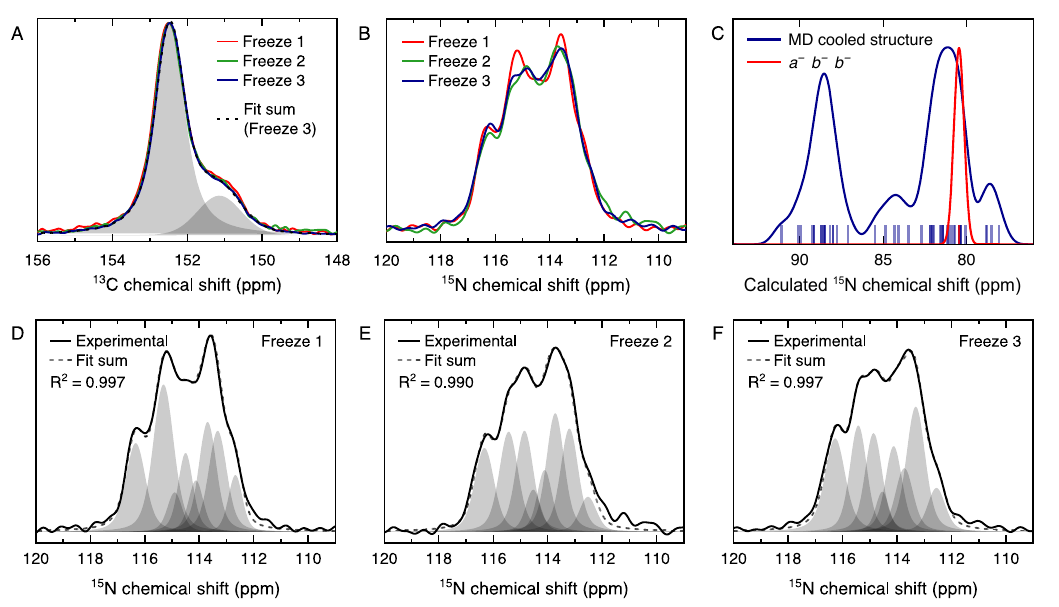}
    \caption{
        Low-temperature $^{1}$H-$^{13}$C (A) and $^{1}$H-$^{15}$N (B) cross polarisation \gls{mas} \gls{nmr} spectra (\qty{9.4}{\tesla}, \qty{8}{\kilo\hertz}) of 3D \ce{FAPbI3} single crystals acquired at \qty{95}{\kelvin} during three sequential freeze-thaw cycles. (C) Calculated $^{15}$N chemical shift distribution for 3D \ce{FAPbI3} at \qty{95}{\kelvin}. (D-F) Fitting of $^{15}$N spectra from each freeze event demonstrating that overall spectra are the cumulative result of varying the population of eight distinct signals.
    }
    \label{fig:nmr-experiments}
\end{figure*}

To further validate the low-temperature phase found in simulations, we compare our results with nuclear magnetic resonance (NMR) spectroscopy and inelastic neutron scattering (INS) experiments at \qty{95}{\kelvin} and \qty{10}{\kelvin}, respectively. The NMR spectra provide insight into the local environment of FA in the $\gamma$-phase, revealing structural changes upon repeated freeze-thaw cycles. While the $^{13}$C spectra are identical for each of the three freeze-thaw cycles (\autoref{fig:nmr-experiments}A), the $^{15}$N spectra show a distribution of several overlapping signals with slight differences in their relative population between each cycle (\autoref{fig:nmr-experiments}B), suggesting that the local structure can change in each freezing event.
To better understand the origin and variability of the $^{15}$N lineshape, we perform chemical shield calculations (\autoref{fig:nmr-experiments}C). These calculations are carried out on the ground state $a^-b^-b^-$ structure and the cooled $a^-a^-c^+$ structure. In the ordered $a^-b^-b^-$ phase, all N atoms are equivalent, resulting in a single chemical shift value. In contrast, the disordered cooled structure exhibits a broad distribution of $^{15}$N chemical shifts due to variations in the local environment. We note that our calculations systematically underestimate the absolute chemical shift values compared to the experiment, which is expected as they do not include spin-orbit coupling effects. Additionallty, since they are performed in rather small supercells, they do not exactly reflect the distribution of FA orientations. Nevertheless, our calculations qualitatively demonstrate that the experimentally observed distribution of $^{15}$N chemical shifts can only be explained by cation disorder, as found in the cooled structure. This also allows us to again rule out the ordered ground state structure as the one present in the experiments.

Fitting of the experimental $^{15}$N low-temperature \gls{mas} \gls{nmr} data shows eight distinct sites, whose  relative populations vary from cycle to cycle (\autoref{fig:nmr-experiments}D--F).
This small number of well-defined FA local environments at \qty{95}{\kelvin} is consistent with the result of our \gls{md} run where we found that, in the low temperature phase, the N--N and C--H vectors point in a limited number of directions.

\begin{figure}
    \centering
    \includegraphics{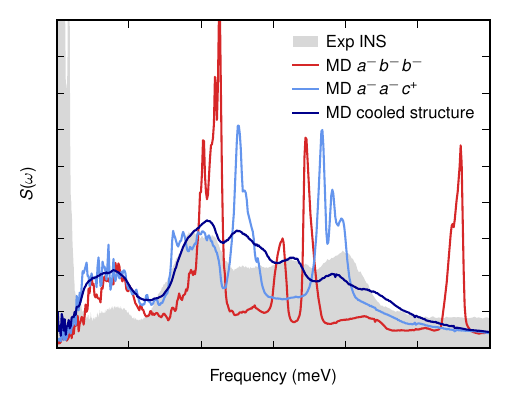}
    \caption{
        Simulated inelastic neutron scattering spectra, $S(\omega)$, for three different structures compared to the experimental spectra from Ref.~\citenum{druzbicki_cation_2021} at \qty{10}{\kelvin}.
        Here, $a^-b^-b^-$ and $a^-a^-c^+$ refers to the structures found from the ground-state search carried out in \autoref{fig:ground_state}, and cooled structure refers to the structure found upon cooling.
       The experimental and simulated spectra are scaled with an arbitrary constant to make them appear on the same scale.
    }
    \label{fig:ins}
\end{figure}

We next compare the $a^-a^-c^+$ structure, identified in our simulations as the best representation of the low-temperature $\gamma$ phase of \ce{FAPbI3}, with the experimental data previously reported in the literature.
Single-crystal X-ray diffraction on the $\gamma$ phase has been challenging since around \qty{100}{\kelvin} the Bragg peaks substantially broaden and split, leading to many unindexed reflections and preventing structure refinement \cite{StoMalKan13}.
On the other hand, structural information on this low-temperature phase can also be accessed through the vibrational signatures of FA obtained in inelastic neutron scattering experiments \cite{druzbicki_cation_2021}.
Here, we compare the vibrational spectra computed for our $a^-b^-b^-$, ideal $a^-a^-c^+$, and $a^-a^-c^+$ structures obtained from \gls{md} runs to the experimental data (\autoref{fig:ins}).

We compute the dynamical structure factor, which is dominated by hydrogen motion due to its large incoherent scattering length.
We find that the spectra obtained from \gls{md} simulations starting from the ideal $a^-b^-b^-$ and $a^-a^-c^+$ structures contain sharp peaks, whereas the spectrum for the structure found upon cooling is substantially broader and agrees well with the experimental spectrum.
This is likely due to the uniform FA ordering and local environments in the ideal structures, which result in sharp peaks, whereas the more disordered cooled structure exhibits a broader spectrum due to the presence of multiple distinct hydrogen environments. This agreement suggests that the $a^-a^-c^+$ structure obtained from the cooling run closely resembles the experimentally observed low-temperature $\gamma$ phase.
Therefore, we conclude that our model likely provides an accurate atomic-level description of the disordered $\gamma$ phase.

\section{Discussion}

The insights gained from our analysis shed light on the low-temperature phase of \ce{FAPbI3}.
\Gls{md} simulations are inherently limited by timescale and computational constraints, often resulting in faster cooling rates and inadequate sampling of the energy landscape.
This limitation frequently leads to kinetic trapping in local minima.
For instance, while the true \gls{gs} of \ce{MAPbI3} is the orthorhombic $a^-a^-c^+$ phase, cooling \gls{md} simulations result in the intermediate tetragonal $a^0a^0c^-$ phase persisting down to \qty{0}{\kelvin}.
Similarly, in \ce{FAPbI3}, the $a^-a^-c^+$ phase is identified as a local minimum below \qty{120}{\kelvin} in cooling \gls{md} runs.
Notably, existing literature remains inconclusive about the low-temperature phase of \ce{FAPbI3}, highlighting the need for further clarification.
Our study identifies the $a^-b^-b^-$ phase as the \gls{gs} and investigates the factors that might prevent the system from transitioning to this state.
Specifically, we analyze two components that can influence the system's behavior: (i) the inorganic framework, focusing on tilting patterns, and (ii) the organic framework, analyzing the orientation preferences and rotational dynamics of FA molecules.

The transition from the tetragonal $a^0a^0c^+$ phase to the orthorhombic 
$a^-b^-b^-$ \gls{gs} requires switching from in-phase to out-of-phase tilts relative to the $c$-axis.
This transition involves an energy barrier that likely stabilizes the 
$a^-a^-c^+$ phase by preserving the in-phase tilt along the $c$-direction.
Focusing on the organic part of the system, the FA molecules exhibit distinct behavior in different phases.
In the \gls{gs} $a^-b^-b^-$  phase, the FA molecules are highly ordered, as evidenced by sharp peaks in the simulated inelastic neutron spectra (\autoref{fig:ins}). The cooled $a^-a^-c^+$ phase exhibits significant disorder, also reflected in the broader peaks in its NMR spectra (\autoref{fig:nmr-experiments}), which closely resemble experimental results and align with the observed structure in experiments.

Moreover, transitioning from the 
$a^-a^-c^+$ phase to the \gls{gs} requires the FA molecules to overcome an additional energy barrier exceeding \qty{100}{\milli\electronvolt\per\atom} to adopt the ordered orientation of the \gls{gs} phase.
This observation is corroborated by the extended rotational relaxation times of FA molecules at lower temperatures (\autoref{fig:arrhenius}), indicating a ``freezing'' effect.

Thus, the freezing of FA molecules appears to be an intrinsic feature of \ce{FAPbI3}, locking the system in the metastable $a^-a^-c^+$ phase.
This phenomenon might explain some of the uncertainties in experimental studies of \ce{FAPbI3} and highlights the local structural variability and the complexity of its underlying dynamics.

\section{Conclusions}

In conclusion, we use an interatomic machine-learned potential to investigate phase transitions and the dynamics of FA cations, aiming to clarify the low-temperature phase of \ce{FAPbI3}.
Our simulations show good agreement with experiments, reproducing the two successive phase transitions: from cubic  $a^0a^0a^0$ to tetragonal $a^0a^0c^+$, and from tetragonal $a^0a^0c^+$ to the low-temperature phase, which we identified as the $a^-a^-c^+$ phase.
Furthermore, our structural search identifies the global \gls{gs} as the orthorhombic $a^-b^-b^-$ phase.

As previously noted, our simulations do not reach the \gls{gs} $a^-b^-b^-$ phase but instead, become kinetically trapped in a local minimum, consistent with findings from existing experimental studies.
This trapping can be attributed to the preferred orientations of FA molecules, which create a complex energy landscape with numerous shallow local minima.
Our analysis suggests that in the $a^-a^-c^+$ phase, FA molecules are effectively ``frozen'', exhibiting very slow rotational dynamics—a behavior also observed in experiments.

Additionally, the transition from the disordered $a^-a^-c^+$ phase to the \gls{gs} $a^-b^-b^-$ phase involves significant energy barriers.
These barriers arise from the need to switch the inorganic tilting pattern from in-phase to out-of-phase and to achieve the highly ordered orientation of FA molecules in the \gls{gs} phase.
Overcoming these barriers is particularly challenging for the disordered 
$a^-a^-c^+$phase.

We believe that this work provides new insights into the low-temperature phase of \ce{FAPbI3}, offering a detailed explanation of FA dynamics and the factors influencing kinetic trapping.
Our findings help resolve existing ambiguities in the literature and advance our understanding of the structural and dynamic complexities of this material.

\section{Acknowledgments}

Funding from the Swedish Strategic Research Foundation through a Future Research Leader programme (FFL21-0129), the Swedish Energy Agency (grant No. 45410-1), the Swedish Research Council (2018-06482, 2019-03993, and 2020-04935), the European Research Council (ERC Starting Grant No. 101162195), the Knut and Alice Wallenberg Foundation (Nos.~2023.0032 and 2024.0042), and the Area of Advance Nano at Chalmers is gratefully acknowledged. 
The computations were enabled by resources provided by the National Academic Infrastructure for Supercomputing in Sweden (NAISS) at C3SE, PDC, and NSC, partially funded by the Swedish Research Council through grant agreement no. 2022-06725.

B.M.G. and D.J.K. acknowledge the UKRI Horizon Europe guarantee funding (PhotoPeroNMR, Grant Agreement EP/Y01376X/1) and the European Union's Horizon 2020 research and innovation programme under grant agreement No 101008500 (PANACEA) for access to low-temperature MAS NMR infrastructure at the University of Gothenburg.

We thank Rasmus Lavén and Maths Karlsson for providing us with their inelastic neutron scattering data for \ce{FAPbI3} and fruitful discussions.

\section{Data availability statement}

The raw NMR and XRD data are available on Zenodo:
[link to be added at proof stage]

\end{document}